\documentclass[prl,aps,showpacs,twocolumn]{revtex4}

\usepackage{epsfig}
\usepackage{epsf}
\usepackage{bm}                 
\usepackage{amsmath}
\usepackage{amssymb}
\usepackage{dcolumn}
\usepackage{graphicx}
\usepackage{psfrag}

\def\ra{\rangle}
\def\la{\langle}
\def\up{\text{\mbox{$\uparrow$}}}
\def\down{\text{\mbox{$\downarrow$}}}

\begin{document}


\title{Efficient high-fidelity quantum computation using matter qubits
  and linear optics}

\author{Sean D.\ Barrett} \email{sean.barrett@hp.com}

\author{Pieter Kok} \email{pieter.kok@hp.com}

\affiliation{Hewlett Packard Laboratories, Filton Road, Stoke
Gifford, Bristol BS34 8QZ, UK}

\date{\today}

\begin{abstract}
 We propose a practical, scalable, and efficient scheme for
 quantum computation using spatially separated matter qubits and
 single photon interference effects. The qubit systems can be
 NV-centers in diamond, Pauli-blockade quantum dots with an excess
 electron or trapped ions with optical transitions, which are each
 placed in a cavity and subsequently entangled using a
 double-heralded single-photon detection scheme. The fidelity of
 the resulting entanglement is extremely robust against the most
 important errors such as detector loss, spontaneous emission, and
 mismatch of cavity parameters. We demonstrate how this entangling
 operation can be used to efficiently generate \emph{cluster
 states} of many qubits, which, together with single qubit
 operations and readout, can be used to implement universal
 quantum computation. Existing experimental parameters indicate
 that high fidelity clusters can be generated with a moderate
 constant overhead.
\end{abstract}


\maketitle

Quantum computation (QC) offers a potentially exponential
computational speed-up over classical computers, and many physical
implementations have been proposed. Particularly promising
proposals are those in which unitary operations and readout in
matter qubits are implemented via laser-driven optical
transitions. Examples are the original ion-trap proposal
\cite{cirac95}, NV-centers in diamond \cite{JelezkoPRL2004}, and
schemes utilizing the Pauli-blockade effect in quantum dots with a
single excess electron \cite{Pazy2003,NazirSpin2004}. Single qubit
operations and readout, using a combination of optical and RF
control fields, have already been demonstrated in ion trap and
NV-diamond systems
\cite{TeleportationInnsbruck,TeleportationNIST,JelezkoPRL2004},
while a number of promising techniques for optically addressing
quantum dot spin qubits have been proposed
\cite{Pazy2003,NazirSpin2004}. In all these cases, the ratio of
the single qubit operation time to the intrinsic decoherence times
suggests that very high fidelity operations are possible.

However, there are substantial difficulties in \emph{scaling}
these implementations to the large numbers of qubits required for
useful QC. Multi-qubit gates are facilitated by a direct
interaction between qubits. Thus adding a new qubit to a quantum
register, together with the associated control fields, necessarily
modifies the Hamiltonian of the system. This can mean that, as
more qubits are added, logic gate implementations become
progressively more complex, and furthermore, new decoherence
channels can be introduced. Furthermore, the need to optically
address individual qubits (e.g. in NV diamond or quantum dot
systems) can lead to seemingly contradictory system requirements:
the qubits need to be sufficiently well separated to be resolved
by the optical field, but must be close enough such that two-qubit
logic can be implemented via the inter-qubit interaction.

\begin{figure}[t]
  \begin{center}
  \begin{psfrags}
     \psfrag{a}{$|\up\ra$}
     \psfrag{b}{$|\down\ra$}
     \psfrag{c}{$|e\ra$}
     \psfrag{d}{$\pi$}
     \psfrag{e}{$g$}
     \psfrag{bs}{BS}
     \psfrag{d1}{$D_+$}
     \psfrag{d2}{$D_-$}
     \psfrag{p}[r]{$\pi$-pulse}
       \epsfig{file=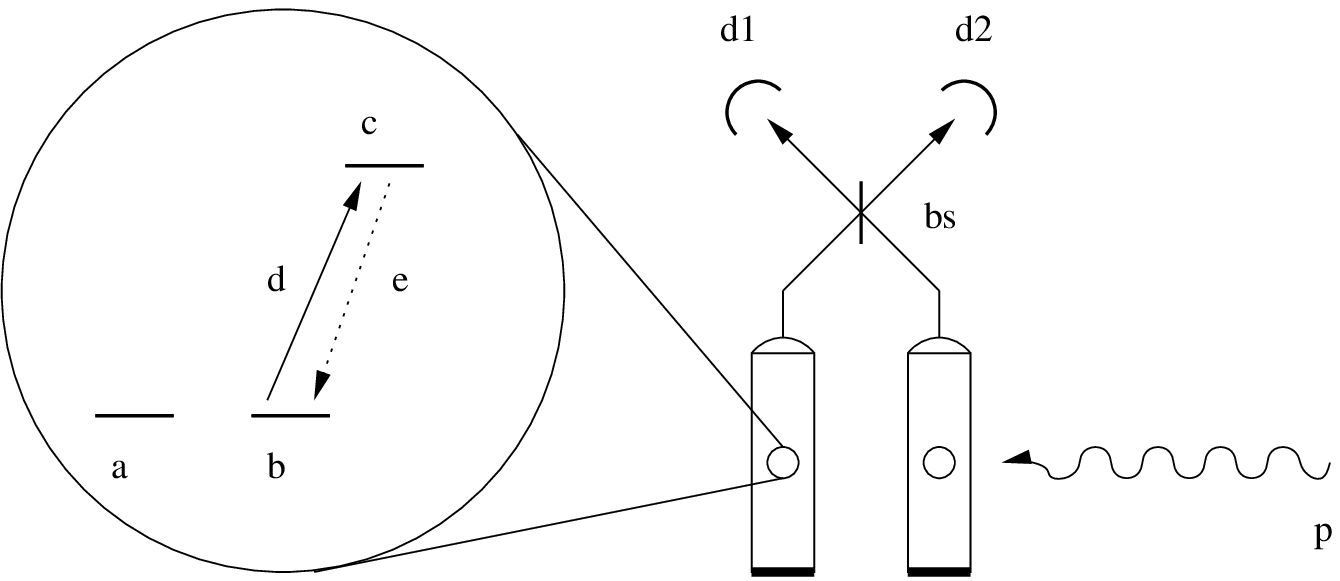, height=3cm}
  \end{psfrags}
  \end{center}
  \caption{In the circle: The qubit system $\{ |\up\ra,|\down\ra\}$
  with the excited state $|e\ra$. The $\pi$-pulse affects only the
  transition $|\down\ra \to |e\ra$, and the emission of a
  photon into the cavity mode brings the excited state back to the qubit
  state $|\down\ra$. Using a pair of leaky cavities with this system
  and conditioning on a single-photon in the detectors $D_+$ or
  $D_-$ creates entangled qubit states. The 50:50 beam splitter (BS)
  is used to erase the which-path information.}
  \label{fig1}
\end{figure}

A potentially promising solution to these scaling challenges is to
perform \emph{distributed} qunatum computing, in which the matter
qubits are spatially separated. In this case, there is no direct
interaction between the qubits. Instead, Entangling Operations
(EOs) between qubits are implemented via single photon
interference effects. A number of schemes to entangle pairs of
distant qubits in this way have been proposed
\cite{cabrillo99,bose99,Feng2003,Duan2003,Browne2003,Simon2003}.
Recently, it has been shown that unitary logic gates can be can be
performed in this manner \cite{Protsenko,Zou}. However, the latter
schemes are either inherently non-deterministic \cite{Zou} or are
sensitive to photon loss or photodetector inefficiency
\cite{Protsenko} and it is not clear whether they can be
used for scalable QC. Other schemes using single photon
interference effects together with local {\em two-qubit} unitary
operations have also been proposed \cite{Duan2004,Taylor}.

In this Letter, we propose a \emph{fully scalable} scheme for
distributed QC using individual matter qubits assuming only single
qubit operations. Our scheme is robust to
photon loss and other sources of errors, and uses
optical transitions of the qubit system, together with linear
optics and photodetection to entangle pairs of spatially
seperated matter qubits in a non-deterministic manner. A key
observation is that even such a non-deterministic EO is sufficient
for scalable QC: our EO can be used to efficiently generate
\emph{cluster states} of many qubits, which, together with
single qubit operations and measurements, are
capable of universal QC \cite{rausschendorf01}. In the context of
linear optics QC \cite{knill01}, it has recently been shown
\cite{yoran03,nielsen04,browne04} that the cluster state model can
be used to significantly reduce the resource overheads required
for scalable QC.

We consider matter systems comprised of two long-lived, low
lying states $|\up\ra$ and $|\down\ra$, and one excited state
$|e\ra$, in an $L$-configuration (see Fig.~\ref{fig1}). The system
is constructed in such a way that an optical $\pi$-pulse will
induce the transformation $|\down\ra \rightarrow |e\ra$ and
$|\up\ra \rightarrow |\up\ra$. The transition $|\up\ra
\leftrightarrow |e\ra$ is forbidden, e.g., by a selection rule.
The states $|\up\ra$ and $|\down\ra$ represent the
logical qubit states $|0\ra$ and $|1\ra$ respectively. We assume
that high fidelity single qubit operations and measurements can be
performed on these logical qubits. Physical systems that have a
suitable level structure include NV-centers in diamond
\cite{JelezkoPRL2004}, quantum dots with a single excess electron
\cite{Pazy2003,NazirSpin2004}, and various trapped ion and atomic
systems. Each such system is embedded in a separate optical
cavity, such that only the $|\down\ra \leftrightarrow |e\ra$ transition
is coupled to the cavity mode. One end of each
cavity is leaky, with the leakage rate of the $i^{\mathrm{th}}$
cavity given by $2\kappa_i$. The light escaping from the cavities
is mixed on a 50:50 beam splitter, the output modes of which are
monitored by two vacuum-discriminating detectors, $D_+$ and $D_-$,
with efficiency $\eta$.

The scheme proceeds as follows. Firstly, both qubits are prepared
in the state $|+\ra = \left(|\up\ra + |\down\ra\right)/\sqrt{2}$
using local unitaries. We then implement the following sequence of
operations: {\em i}) Apply an optical $\pi$-pulse to each qubit,
coherently pumping the population in the $|\down\ra$ state into
the $|e\ra$ state; {\em ii}) Wait for up to a time
$t_{\textrm{wait}}$ for a photo-detection event in either $D_+$ or
$D_-$; {\em iii}) Wait for a further time $t_{\textrm{relax}}$ for
any remaining excitation in the qubit-cavity systems to relax;
{\em iv}) Apply an $X$ operation to both qubits, coherently
flipping the spins as $|\up\ra \to |\down\ra$ and $|\down\ra \to
|\up\ra$; {\em v}) Repeat steps {\em i}--{\em iii}.

Appropriate values for $t_{\textrm{wait}}$ and
$t_{\textrm{relax}}$ are determined by the system parameters discussed
below. If zero or two photo-detection events are
observed on either round of the procedure, the scheme failed, and the
qubits must be newly prepared before
re-attempting the entangling procedure. On the other hand, if one
(and only one) photo-detection event is observed on each round of
the protocol, the scheme has succeeded, and a maximally entangled state
is prepared \emph{with unit fidelity} (given ideal systems).
We call this technique {\em double heralding}, and it turns out to
be exceedingly robust against most common experimental errors.

We analyzed the scheme in detail using the quantum trajectories
formalism \cite{carmichael}. For clarity, we first consider the
ideal case, in which the detectors have unit efficiency
($\eta=1$), and spontaneous emission of photons from the
transition $|e\ra \rightarrow |\down\ra$ into modes other than the
cavity mode is neglected. During time periods where no detector
clicks are observed, the conditional state of the system, in the
interaction picture, evolves smoothly according to the effective
Hamiltonian ($\hbar=1$)
\begin{equation}
H_{\textrm{eff}} = \sum_{i=A,B} \frac{g_i}{2} \left( | \down
\ra_i{}_i\la e | \hat{c}^\dag_i + \textrm{H. c.} \right) - i
\sum_{i=A,B} \kappa_i \hat{c}^\dag_i \hat{c}_i \,.
\end{equation}
Here, $g_i$ denotes the Jaynes-Cummings coupling between the
$|e\ra_i \leftrightarrow |\down\ra_i$ transition and the mode of
the $i^{\mathrm th}$ cavity, and $\hat{c}_i$ is the corresponding
annihilation operator. For the purpose of illustrating the ideal
case, we assume that systems $A$ and $B$ are identical, such that
$g_A = g_B=g$ and $\kappa_A = \kappa_B = \kappa$, and that $\kappa
\ge g$.

When a single click is observed in detector $D_\pm$, the state of
the whole system discontinuously evolves as $|\psi(t)\ra
\rightarrow \hat{c}_\pm |\psi(t)\ra$, where $\hat{c}_\pm =
(\hat{c}_A \pm \hat{c}_B)/\sqrt{2}$ denotes the corresponding jump
operators. Thus, after steps {\em i}--{\em ii} of the entangling
protocol, conditioned on observing a detector click at time $t_1
\le t_{\textrm{wait}}$, the unnormalized state of the whole system
is
\begin{eqnarray}
 |\tilde{\psi}(t_1)\ra
 &=& \alpha(t_1) |\Psi_\pm\ra +
 \alpha(t_1)\beta(t_1)\frac{|\down,0;e,0\ra \pm
 |e,0;\down,0\ra}{\sqrt{2}} \cr &&+
 2\alpha^2(t_1)\frac{|\down,0;\down,1\ra \pm
|\down,1;\down,0\ra}{\sqrt{2}} \,. \label{StateAfterClick1}
\end{eqnarray}
Here, $|q_A,p_A;q_B,p_B\ra$ is the state of the whole system,
with $q_{A(B)}$ and $p_{A(B)}$ denoting the states of matter
system $A$ $(B)$ and cavity mode $A$ $(B)$ respectively,
$|\Psi^\pm\ra = (|\down,0;\up,0\ra \pm
|\up,0;\down,0\ra)/\sqrt{2}$ are maximally entangled states,
$\alpha(t) = -ig /(4 \sqrt{\kappa^2 - g^2})
\left(e^{-\Gamma_{\textrm{slow}} t/2} -e^{-\Gamma_{\textrm{fast}}
t/2} \right)$ and $\beta(t) = \frac{1}{2}(1 + \kappa/
\sqrt{\kappa^2-g^2}) e^{-\Gamma_{\textrm{slow}} t/2} +
\frac{1}{2}(1- \kappa/\sqrt{\kappa^2-g^2})
e^{-\Gamma_{\textrm{fast}} t/2}$, where $\Gamma_{\textrm{fast}} =
\kappa+\sqrt{\kappa^2-g^2}$ and $\Gamma_{\textrm{slow}} =
\kappa-\sqrt{\kappa^2-g^2}$. In order to obtain a detector click
with significant probability, $t_{\textrm{wait}}$ should be chosen
to be a few times $\Gamma_{\textrm{slow}}^{-1}$.

Equation (\ref{StateAfterClick1}) implies that it may be possible
to observe a \emph{second} detector click on the \emph{first}
round of the protocol. However, realistic photo-detectors
typically cannot resolve two photons arriving in quick succession
\cite{kok00}. Within the quantum trajectories description, this
can be simulated by assuming that no information is available from
either detector after the first click. After a time
$t_{\textrm{relax}} \gg \Gamma_{\textrm{slow}}^{-1}$, the system
decoheres to the state
\begin{equation}
\rho = \frac{1}{N}|\Psi_\pm\ra\la\Psi_\pm| +
\left(1-\frac{1}{N} \right)| \down \down\ra\la
\down\down | \,, \label{StateAfterFirstRelaxation}
\end{equation}
where $N = 1 +  |\beta(t_1)|^2 + |\alpha(t_1)|^2$. The undesirable
second term in 
Eq.~(\ref{StateAfterFirstRelaxation}) is removed by
applying steps {\em iv} and {\em v} of the entangling procedure.
If a photo-detection occurs on the second round, the final state
of the system is a pure, maximally entangled state. If the two
clicks are observed in the same (different) detector(s), the final
state is $|\Psi^+\ra$ ($|\Psi^-\ra$). Each of the four possible
successful outcomes occurs with probability $1/8$, leading to a
total success probability of $p=\frac{1}{2}$. 

\begin{figure}[t]
  \begin{center}
    \epsfig{file=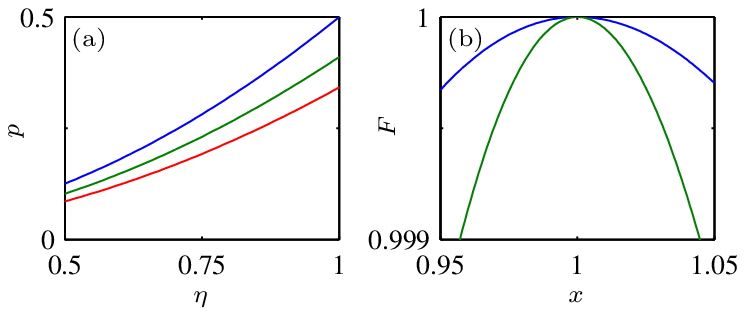}
      \end{center}
  \caption{(a) Probability of success $p$ versus the detector efficiency
    $\eta$ after both rounds of the entangling procedure. Each curve
    is plotted for a different value of the spontaneous emission rate
    $\gamma=\{0, 0.1 \Gamma_{\textrm{slow}},0.2 \Gamma_{\textrm{slow}}\}$ (top to bottom). The other
    parameters used were $g = 0.3$ and $\kappa = 1$.
(b) Fidelity of the entangled qubit pair for imperfectly
mode-matched photons, calculated using the quantum trajectories
formalism. Upper curve: fidelity for non identical leakage rates,
$x=\kappa_1/\kappa_2$, taking $g_1=g_2$. Lower curve:fidelity for
non identical coupling parameters, $x=g_1/g_2$, taking $g_2=0.3$
and $\kappa_1 = \kappa_2 = 1$.} \label{p_succ}
\end{figure}

We also analyzed the scheme in the non-ideal case, allowing for
less than perfect detector efficiency $\eta<1$, and finite
spontaneous emission into free space ($\gamma_1 = \gamma_2 =
\gamma > 0$). These imperfections do not reduce the fidelity of
the final state, but do reduce the success probability (see
Fig.~\ref{p_succ}a). Note $p$ has a quadratic dependence on
$\eta$, while $p$ decreases rapidly for $\gamma \gtrsim
\Gamma_{\textrm{slow}}$.

The dominating experimental imperfections that do reduce the fidelity
can be classified into three groups: (1) decoherence of the matter
qubits; (2) dark counts in the detectors; and (3) imperfect  mode
matching of the photons incident on the beam splitter. Firstly, the
effect of spin decoherence depends on the way the cluster states are
generated, and can be estimated by comparing the spin decoherence time
$t_{\textrm{d}}$ with the ``clock time'' $t_{\textrm{c}} \sim 10
\Gamma^{-1}_{\textrm{slow}}$ at which the EO can be repeated. If the
preparation of cluster states is performed in parallel, the typical
time overhead is $m$ clock cycles (see below). Thus the
average age of a qubit the moment it is added to the cluster is
$(m/2)t_{\textrm{c}}$, and $m \lesssim 8$ for reasonable detector
efficiencies. Assuming a reasonably strong cavity qubit coupling,
$g=100 \gamma$, and critically damped cavities ($g\approx\kappa$), the
size of errors due to spin decoherence is given by $\varepsilon \sim
(m/2)t_{\textrm{c}}/t_{\textrm{d}} \sim 0.4 \gamma^{-1} /
t_{\textrm{d}}$. For instance, for the NV-diamond system
($\gamma^{-1}=25$ ns \cite{beveratos2001} and $t_{\textrm{d}} = 32$
$\mu$s \cite{Kennedy2002}), we have $\varepsilon \sim 3 \times
10^{-4}$.

Secondly, detector dark counts on either round of the EO can lead to a
spurious `success' of the EO, which can reduce the fidelity of the
entanglement. For existing APD detectors, dark count rates are
typically $\Gamma_{\textrm{dc}} < 500$ s$^{-1}$
\cite{PerkinElmerDataSheet}. Dark counts can be made negligible by
observing the detector output only for the window $t_{\textrm{wait}}
\sim 3 \Gamma^{-1}_{\textrm{slow}}$ ($\sim 1$ ns for NV-diamond). The
probability of a spurious count is therefore
$p_{\textrm{dc}}=\Gamma_{\textrm{dc}}t_{\textrm{wait}} \sim
10^{-7}$. Thus dark counts should have a negligible effect on the
cluster fidelity.

Finally, imperfect mode matching of the photons emitted by the matter
qubit-cavity systems reduces the fidelity, because the photons carry
information regarding their origin. Non identical central frequencies,
different polarizations, and spatio-temporal mode shapes of the
photons can all reduce the fidelity.  The frequency of the photons
emitted from cavity $i$ depends on the frequencies of both the
$|\down\ra_i \leftrightarrow |e\ra_i$ transition ($\omega_{\down e,
i}$) and the cavity mode ($\omega_{\textrm{cav}, i}$). The
$\omega_{\down e, i}$'s can be tuned independently, e.g. by using
local electric and magnetic fields to induce Stark and Zeeman
shifts. The $\omega_{\textrm{cav}, i}$'s can also be accurately and
independently tuned, e.g. by using strain-tunable silica microcavities
\cite{vonKlitzing2001}, or piezoelectrically tuned fibre optic
microcavities \cite{JasonSmithPrivateCommunication}.  The polarization
of the emitted photons can be accurately matched using linear optical
elements \cite{BachorTextbook}. The spatio-temporal mode shapes of the
emitted photons depend on the $g_i$'s and $\kappa_i$'s of the
respective cavities. These parameters depend on the structure of the
cavities, and hence are more difficult to calibrate once the cavities
have been fabricated. However, we calculated that the EO is rather
robust to mismatches in the $g_i$'s and $\kappa_i$'s (see Fig
\ref{p_succ}b): mismatches of a few percent lead to a reduction in
fidelity of less than $10^{-3}$. Deterministic sources of
indistinguishable photons, which have similar requirements to those
needed for our scheme, are currently being developed by a number of
groups
\cite{Pelton2002,Stace2003,meier2004,beveratos2001,McKeever2004}.

\begin{figure}[t]
  \begin{center}
  \begin{psfrags}
     \psfrag{a}{a)}
     \psfrag{b}{b)}
     \psfrag{c}{$A_2$}
     \psfrag{1}{$A_1$}
     \psfrag{2}{$B_1$}
     \psfrag{3}{$B_2$}
     \psfrag{4}{$B_3$}
        \epsfig{file=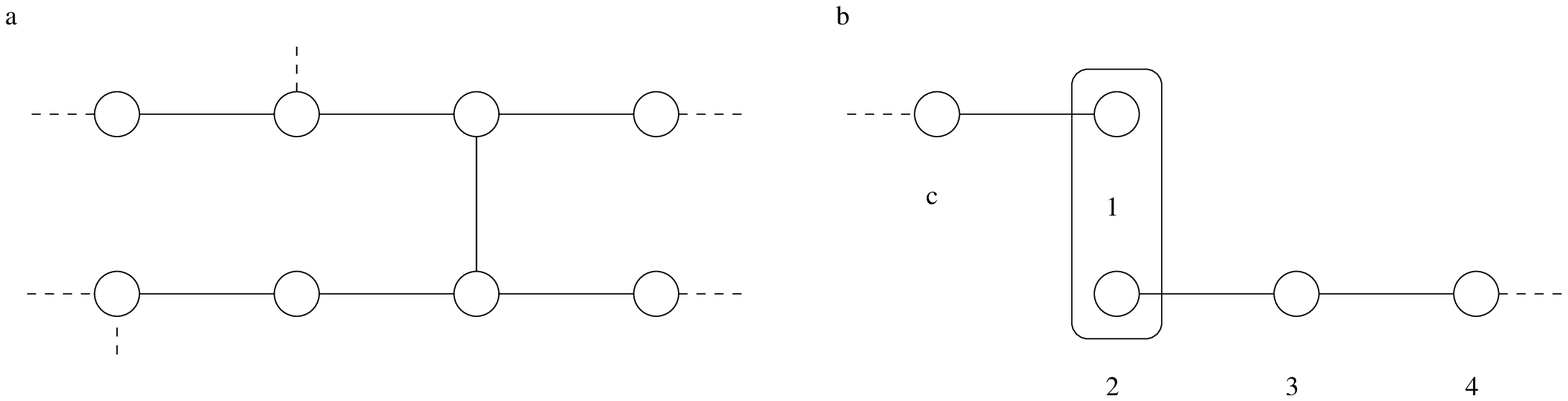,height=2.2cm}
  \end{psfrags}
  \end{center}
  \caption{Cluster states with arbitrary single-qubit measurements can
  perform a universal quantum computation. a) The qubits (circles) are
  entangled with their nearest horizontal neighbor via
  EOs (depicted by lines between the
  qubits), and gates between computational qubits are incorporated by
  the vertical lines. b) Linear clusters can be joined together by applying
  the EO between qubits $A_1$ and $B_1$, and subsequently performing single
  qubit operations and measurements.}
  \label{fig2}
\end{figure}

The next step towards scalable quantum computers is linking qubits
together into cluster states, using the EO described above. A
cluster state of qubits $\left\{q_1,q_2,..q_N \right\}$ can be
represented graphically by a collection of qubit nodes connected
by edges connecting neighboring qubits, as depicted in
Fig.~\ref{fig2}a.
A linear cluster of $N$ qubits (a {\em chain}) may be represented
in the form $|C\ra_{1...N}=(|\up\ra_1 + |\down \ra_1 Z_2
)(|\up\ra_2 + |\down \ra_2 Z_3 )...(|\up\ra_N + |\down \ra_N )$,
where $Z_i$ represents the Pauli phase-flip operation acting on
qubit $i$. Such linear clusters can be grown using our EO, as we
now describe. Given a cluster $|C\ra_{2...N}$, qubit 1 can be
added to the end of the cluster by first preparing qubit 1 in the
state $|+\ra_1 \equiv |\down\ra_1 + |\up\ra_1$ and then applying
the EO to qubits 1 and 2. If the EO is successful, the resulting
state is of the form $(|\up\ra_1 |\down\ra_2 \pm |\down \ra_1|\up
\ra_2 Z_3)|C\ra_{3...N}$, depending on whether both clicks were
observed in the same detector. This can be transformed into a
cluster state by applying the local operations $H_1 X_2$ or $X_1
H_1 X_2$, conditional on the outcome of the EO (here $H_i$ is the
Hadamard operation, and $X_i$ the Pauli operator implementing a
bit flip). If the EO fails, the state of qubit 2 is, in general,
unknown. However, measuring qubit 2 in the computational basis
removes qubit 2 from the cluster, but projects qubits $\{3 \ldots
N\}$ back into a pure cluster state. Therefore, failure of the EO
causes the original cluster to shrink by 1 qubit.

Repeatedly applying the procedure described above allows long
chains to be grown. However, the theoretical upper limit on the
success probability of our protocol is $p=\frac{1}{2}$, and when
the protocol fails, the chain shrinks by 1 qubit. Therefore, with
only the procedure described above one can not create large
clusters efficiently. If recycling of the clusters after a failure
\cite{browne04} is not performed, the average number of EOs
required to create a cluster of $m$ qubits is $N_{\mathrm{EO}} =
\sum_{i=1}^{m-1} p^{-i} = p^{1-m}(1-p^{m-2})/(1-p)$.

A way around this problem is a `divide and conquer' approach, in
which short chains are grown inefficiently, and then joined to a
longer cluster using the EO together with local operations. The EO
can be used to join two clusters, as shown in Fig.~\ref{fig2}b.
The initial state of two chains, $A$ and $B$ of length $N$ and $m$
respectively, may be written $(|C\ra_{\{A\}} |\up\ra_{A_1} +
Z_{A_2}|C\ra_{\{A\}} |\down\ra_{A_1})(|\up\ra_{B_1} |C\ra_{\{B\}} 
+ Z_{B_2} |\down\ra_{B_1} |C\ra_{\{B\}})$, where $|C\ra_{\{A\}}$
($|C\ra_{\{B\}}$) represents a linear cluster state of qubits
$\{A_2 \ldots A_N\}$ ($\{B_2 \ldots B_m\}$). These chains can be
joined together by first performing the local operation $X_{A_1}$,
then applying the EO between qubits $A_1$ and $B_1$, and measuring
$B_1$ in the basis $|\pm\ra_{B_1} \equiv |\down\ra_{B_1} \pm
|\up\ra_{B_1}$. If the EO was successful, the remaining qubits are
left in the state $\pm |C\ra_{\{A\}} |\up\ra_{A_1}|C\ra_{\{B\}}
\pm Z_{A_2}|C\ra_{\{A\}} |\down\ra_{A_1} Z_{B_2}|C\ra_{\{B\}}$,
where the first sign depends on the outcome of the $B_1$
measurement, and the second sign depends on the outcome of the EO.
Applying a local operation to qubit $A_1$ yields a cluster state,
of length $N+m-1$, of qubits $\{A_N, \ldots A_1, B_2, \ldots
B_m\}$. If the EO fails, qubit $A_1$ must be measured in the
computational basis, and the original cluster state shrinks by 1
qubit. Thus the average length of the new cluster is $L = p(N+m-1)
+ (1-p) (N-1)$. In order that the cluster grows on average, we
require $L>N$, which implies that length of the short chains
should satisfy $m>1/p$.

Chains of fixed length $m$ can be grown independently using the
EO, either by sequentially adding single qubits to the end of a
cluster, or by joining sub-chains together. Growing these
$m$-chains adds a constant overhead cost to the cluster generation
process. For example, growing a 4-chain (without recycling)
requires on average $p^{-3} + p^{-2} + p^{-1}$ applications of the
EO, and each attempt to join such a chain adds on average $4p-1$
qubits to the large cluster, leading to a total cost of $C_4 =
(p^{-3} + p^{-2} + p^{-1} + 1)/(4p-1)$ EOs per qubit added to the
large cluster. A 5-chain can be grown by joining two 3-chains
together. Joining such 5-chains to a longer cluster leads to a
total cost of $C_5 = (2p^{-3} + 2p^{-2} + p^{-1} + 1)/(5p-1)$ EOs
per qubit. To minimize these costs, the collection and detection
efficiencies should be maximized. For example, for $p\approx 0.24$
(or $\eta = 70\%$ with $\gamma=0$), we require $m=5$, and we find
$C_5 = 775$. A modest improvement in detector efficiency
dramatically reduces the overhead cost: for $\eta=85\%$ and
$\gamma=0$, we find $C_4 = 73.4$.
%
%
Note that there may be more efficient schemes for growing linear
clusters using our EO (e.g. employing recycling of small clusters
\cite{browne04}) which yield lower overhead costs.

In order to build linear chains into two-dimensional cluster
states capable of simulating arbitrary logic networks, cross links
between linear chains must be constructed \cite{nielsen04}. Such a
link can be created by first using the EO to create an $I$ shaped
cluster (see Fig \ref{fig2}(a)) offline, for some fixed cost.
Provided the arms of this $I$-cluster are sufficiently long, the
EO can be used to join the $I$-cluster to a pair of linear
clusters with a high probability, and therefore create a cross
link between the clusters. This leads to a constant overhead cost
per cross link added to the cluster. Note that other methods for
creating two-dimensional clusters, e.g. using microclusters
\cite{nielsen04} or redundant encoding \cite{browne04}, have also
been proposed.

Our proposal has a number of very desirable features with respect to
practical implementations. Firstly, our scheme requires only a simple
level structure and single-qubit operations. Secondly, photon loss
does not reduce the fidelity of the entangled state of the qubits, but
merely adds to the constant overhead cost. Thirdly, owing to the
simplicity of the optical networks used in this scheme, mode matching
should be relatively straightforward. Fourthly, the scheme is
inherently \emph{distributed}: individual qubit-cavity systems can be
placed in distant labs, and connected by optical fibers. This means
that our scheme lends itself naturally to distributed applications,
such as quantum repeaters \cite{Briegel98} and quantum cryptography
\cite{ekert91}. Finally, many of the techniques described here have
been demonstrated experimentally, and the system requirements needed
to create high-fidelity cluster states do not seem prohibitively
restrictive.

While preparing this manuscript, we became aware of an
alternative scheme that may also be used for generating cluster
states of matter qubits \cite{Lim2004}. We thank Tim Spiller and Bill
Munro for valuable discussions and careful reading of the
manuscript. The authors are supported by the E.U. Nanomagiq and Ramboq
projects.

\vspace{-3mm}

\bibliography{cs3}

\begin{thebibliography}{36}
\expandafter\ifx\csname natexlab\endcsname\relax\def\natexlab#1{#1}\fi
\expandafter\ifx\csname bibnamefont\endcsname\relax
  \def\bibnamefont#1{#1}\fi
\expandafter\ifx\csname bibfnamefont\endcsname\relax
  \def\bibfnamefont#1{#1}\fi
\expandafter\ifx\csname citenamefont\endcsname\relax
  \def\citenamefont#1{#1}\fi
\expandafter\ifx\csname url\endcsname\relax
  \def\url#1{\texttt{#1}}\fi
\expandafter\ifx\csname urlprefix\endcsname\relax\def\urlprefix{URL }\fi
\providecommand{\bibinfo}[2]{#2}
\providecommand{\eprint}[2][]{\url{#2}}

\bibitem[{\citenamefont{Cirac and Zoller}(1995)}]{cirac95}
\bibinfo{author}{\bibfnamefont{J.}~\bibnamefont{Cirac}} \bibnamefont{and}
  \bibinfo{author}{\bibfnamefont{P.}~\bibnamefont{Zoller}},
  \bibinfo{journal}{Phys.\ Rev.\ Lett.} \textbf{\bibinfo{volume}{74}},
  \bibinfo{pages}{4091} (\bibinfo{year}{1995}).

\bibitem[{\citenamefont{Jelezko et~al.}(2004)}]{JelezkoPRL2004}
\bibinfo{author}{\bibfnamefont{F.}~\bibnamefont{Jelezko}} \bibnamefont{et~al.},
  \bibinfo{journal}{Phys. Rev. Lett.} \textbf{\bibinfo{volume}{92}},
  \bibinfo{pages}{076401} (\bibinfo{year}{2004}).

\bibitem[{\citenamefont{Pazy et~al.}(2003)}]{Pazy2003}
\bibinfo{author}{\bibfnamefont{E.}~\bibnamefont{Pazy}} \bibnamefont{et~al.},
  \bibinfo{journal}{Europhys. Lett.} \textbf{\bibinfo{volume}{62}},
  \bibinfo{pages}{175} (\bibinfo{year}{2003}).

\bibitem[{\citenamefont{Nazir et~al.}(2004)}]{NazirSpin2004}
\bibinfo{author}{\bibfnamefont{A.}~\bibnamefont{Nazir}} \bibnamefont{et~al.},
  \bibinfo{journal}{quant-ph/0403225}  (\bibinfo{year}{2004}).

\bibitem[{\citenamefont{Riebe et~al.}(2004)}]{TeleportationInnsbruck}
\bibinfo{author}{\bibfnamefont{M.}~\bibnamefont{Riebe}} \bibnamefont{et~al.},
  \bibinfo{journal}{Nature} \textbf{\bibinfo{volume}{429}},
  \bibinfo{pages}{734} (\bibinfo{year}{2004}).

\bibitem[{\citenamefont{Barrett et~al.}(2004)}]{TeleportationNIST}
\bibinfo{author}{\bibfnamefont{M.~D.} \bibnamefont{Barrett}}
  \bibnamefont{et~al.}, \bibinfo{journal}{Nature}
  \textbf{\bibinfo{volume}{429}}, \bibinfo{pages}{737} (\bibinfo{year}{2004}).

\bibitem[{\citenamefont{Cabrillo et~al.}(1999)}]{cabrillo99}
\bibinfo{author}{\bibfnamefont{C.}~\bibnamefont{Cabrillo}}
  \bibnamefont{et~al.}, \bibinfo{journal}{Phys.\ Rev.\ A}
  \textbf{\bibinfo{volume}{59}}, \bibinfo{pages}{1025} (\bibinfo{year}{1999}).

\bibitem[{\citenamefont{Bose et~al.}(1999)}]{bose99}
\bibinfo{author}{\bibfnamefont{S.}~\bibnamefont{Bose}} \bibnamefont{et~al.},
  \bibinfo{journal}{Phys. Rev. Lett.} \textbf{\bibinfo{volume}{83}},
  \bibinfo{pages}{5158} (\bibinfo{year}{1999}).

\bibitem[{\citenamefont{Feng et~al.}(2003)}]{Feng2003}
\bibinfo{author}{\bibfnamefont{X.~L.} \bibnamefont{Feng}} \bibnamefont{et~al.},
  \bibinfo{journal}{Phys. Rev. Lett.} \textbf{\bibinfo{volume}{90}},
  \bibinfo{pages}{217902} (\bibinfo{year}{2003}).

\bibitem[{\citenamefont{Duan and Kimble}(2003)}]{Duan2003}
\bibinfo{author}{\bibfnamefont{L.-M.} \bibnamefont{Duan}} \bibnamefont{and}
  \bibinfo{author}{\bibfnamefont{H.~J.} \bibnamefont{Kimble}},
  \bibinfo{journal}{Phys. Rev. Lett.} \textbf{\bibinfo{volume}{90}},
  \bibinfo{pages}{253601} (\bibinfo{year}{2003}).

\bibitem[{\citenamefont{Browne et~al.}(2003)\citenamefont{Browne, Plenio, and
  Huelga}}]{Browne2003}
\bibinfo{author}{\bibfnamefont{D.}~\bibnamefont{Browne}},
  \bibinfo{author}{\bibfnamefont{M.}~\bibnamefont{Plenio}}, \bibnamefont{and}
  \bibinfo{author}{\bibfnamefont{S.}~\bibnamefont{Huelga}},
  \bibinfo{journal}{Phys. Rev. Lett.} \textbf{\bibinfo{volume}{91}},
  \bibinfo{pages}{067901} (\bibinfo{year}{2003}).

\bibitem[{\citenamefont{Simon and Irvine}(2003)}]{Simon2003}
\bibinfo{author}{\bibfnamefont{C.}~\bibnamefont{Simon}} \bibnamefont{and}
  \bibinfo{author}{\bibfnamefont{W.~T.~M.} \bibnamefont{Irvine}},
  \bibinfo{journal}{Phys. Rev. Lett.} \textbf{\bibinfo{volume}{91}},
  \bibinfo{pages}{110405} (\bibinfo{year}{2003}).

\bibitem[{\citenamefont{Protsenko et~al.}(2002)}]{Protsenko}
\bibinfo{author}{\bibfnamefont{I.~E.} \bibnamefont{Protsenko}}
  \bibnamefont{et~al.}, \bibinfo{journal}{Phy. Rev. A}
  \textbf{\bibinfo{volume}{66}}, \bibinfo{pages}{062306}
  (\bibinfo{year}{2002}).

\bibitem[{\citenamefont{amd W.~Mathis}(2004)}]{Zou}
\bibinfo{author}{\bibfnamefont{X.~Z.} \bibnamefont{amd W.~Mathis}},
  \bibinfo{journal}{quant-ph/0401042}  (\bibinfo{year}{2004}).

\bibitem[{\citenamefont{Duan et~al.}(2004)}]{Duan2004}
\bibinfo{author}{\bibfnamefont{L.-M.} \bibnamefont{Duan}} \bibnamefont{et~al.},
  \bibinfo{journal}{Quant. Inf. Comp.} \textbf{\bibinfo{volume}{4}},
  \bibinfo{pages}{165} (\bibinfo{year}{2004}).

\bibitem[{\citenamefont{Taylor et~al.}(2004)}]{Taylor}
\bibinfo{author}{\bibfnamefont{J.~M.} \bibnamefont{Taylor}}
  \bibnamefont{et~al.}, \bibinfo{journal}{cond-mat/0407640}
  (\bibinfo{year}{2004}).

\bibitem[{\citenamefont{Raussendorf and Briegel}(2001)}]{rausschendorf01}
\bibinfo{author}{\bibfnamefont{R.}~\bibnamefont{Raussendorf}} \bibnamefont{and}
  \bibinfo{author}{\bibfnamefont{H.}~\bibnamefont{Briegel}},
  \bibinfo{journal}{Phys. Rev. Lett.} \textbf{\bibinfo{volume}{86}},
  \bibinfo{pages}{5188} (\bibinfo{year}{2001}).

\bibitem[{\citenamefont{Knill et~al.}(2001)\citenamefont{Knill, Laflamme, and
  Milburn}}]{knill01}
\bibinfo{author}{\bibfnamefont{E.}~\bibnamefont{Knill}},
  \bibinfo{author}{\bibfnamefont{R.}~\bibnamefont{Laflamme}}, \bibnamefont{and}
  \bibinfo{author}{\bibfnamefont{G.}~\bibnamefont{Milburn}},
  \bibinfo{journal}{Nature} \textbf{\bibinfo{volume}{409}}, \bibinfo{pages}{26}
  (\bibinfo{year}{2001}).

\bibitem[{\citenamefont{Yoran and Reznik}(2003)}]{yoran03}
\bibinfo{author}{\bibfnamefont{N.}~\bibnamefont{Yoran}} \bibnamefont{and}
  \bibinfo{author}{\bibfnamefont{B.}~\bibnamefont{Reznik}},
  \bibinfo{journal}{Phys. Rev. Lett.} \textbf{\bibinfo{volume}{91}},
  \bibinfo{pages}{037903} (\bibinfo{year}{2003}).

\bibitem[{\citenamefont{Nielsen}(2004)}]{nielsen04}
\bibinfo{author}{\bibfnamefont{M.}~\bibnamefont{Nielsen}},
  \bibinfo{journal}{Phys. Rev. Lett.} p. \bibinfo{pages}{(to appear)}
  (\bibinfo{year}{2004}).

\bibitem[{\citenamefont{Browne and Rudolph}(2004)}]{browne04}
\bibinfo{author}{\bibfnamefont{D.}~\bibnamefont{Browne}} \bibnamefont{and}
  \bibinfo{author}{\bibfnamefont{T.}~\bibnamefont{Rudolph}},
  \bibinfo{journal}{quant-ph/0405157}  (\bibinfo{year}{2004}).

\bibitem[{\citenamefont{Carmichael}(1993)}]{carmichael}
\bibinfo{author}{\bibfnamefont{H.}~\bibnamefont{Carmichael}},
  \emph{\bibinfo{title}{An open systems approach to quantum optics}},
  vol.~\bibinfo{volume}{18} of \emph{\bibinfo{series}{Lecture notes in
  physics}} (\bibinfo{publisher}{Springer, Berlin}, \bibinfo{year}{1993}).

\bibitem[{\citenamefont{Kok and Braunstein}(2000)}]{kok00}
\bibinfo{author}{\bibfnamefont{P.}~\bibnamefont{Kok}} \bibnamefont{and}
  \bibinfo{author}{\bibfnamefont{S.}~\bibnamefont{Braunstein}},
  \bibinfo{journal}{Phys. Rev. A} \textbf{\bibinfo{volume}{61}},
  \bibinfo{pages}{042304} (\bibinfo{year}{2000}).

\bibitem[{\citenamefont{Beveratos et~al.}(2002)}]{beveratos2001}
\bibinfo{author}{\bibfnamefont{A.}~\bibnamefont{Beveratos}}
  \bibnamefont{et~al.}, \bibinfo{journal}{Eur.\ Phys.\ J.\ D}
  \textbf{\bibinfo{volume}{18}}, \bibinfo{pages}{191} (\bibinfo{year}{2002}).

\bibitem[{\citenamefont{Kennedy et~al.}(2002)}]{Kennedy2002}
\bibinfo{author}{\bibfnamefont{T.~A.} \bibnamefont{Kennedy}}
  \bibnamefont{et~al.}, \bibinfo{journal}{Phys. Status Solidi (b)}
  \textbf{\bibinfo{volume}{233}}, \bibinfo{pages}{416} (\bibinfo{year}{2002}).

\bibitem[{Per()}]{PerkinElmerDataSheet}
\bibinfo{note}{SPCM-AQR datasheet at www.perkinelmer.com}.

\bibitem[{\citenamefont{VonKlitzing et~al.}(2001)}]{vonKlitzing2001}
\bibinfo{author}{\bibfnamefont{W.}~\bibnamefont{VonKlitzing}}
  \bibnamefont{et~al.}, \bibinfo{journal}{New J. Phys.}
  \textbf{\bibinfo{volume}{3}}, \bibinfo{pages}{14} (\bibinfo{year}{2001}).

\bibitem[{Jas()}]{JasonSmithPrivateCommunication}
\bibinfo{note}{Jason Smith, presonal communication.}

\bibitem[{\citenamefont{Bachor and Ralph}(2004)}]{BachorTextbook}
\bibinfo{author}{\bibfnamefont{H.-A.} \bibnamefont{Bachor}} \bibnamefont{and}
  \bibinfo{author}{\bibfnamefont{T.~C.} \bibnamefont{Ralph}},
  \emph{\bibinfo{title}{An guide to experiments in quantum optics}}
  (\bibinfo{publisher}{Wiley-VCH, Weinheim}, \bibinfo{year}{2004}).

\bibitem[{\citenamefont{Pelton et~al.}(2002)}]{Pelton2002}
\bibinfo{author}{\bibfnamefont{M.}~\bibnamefont{Pelton}} \bibnamefont{et~al.},
  \bibinfo{journal}{Phys. Rev. Lett.} \textbf{\bibinfo{volume}{89}},
  \bibinfo{pages}{233602} (\bibinfo{year}{2002}).

\bibitem[{\citenamefont{Stace et~al.}(2003)\citenamefont{Stace, Milburn, and
  Barnes}}]{Stace2003}
\bibinfo{author}{\bibfnamefont{T.}~\bibnamefont{Stace}},
  \bibinfo{author}{\bibfnamefont{G.}~\bibnamefont{Milburn}}, \bibnamefont{and}
  \bibinfo{author}{\bibfnamefont{C.}~\bibnamefont{Barnes}},
  \bibinfo{journal}{Phys. Rev. B} \textbf{\bibinfo{volume}{67}},
  \bibinfo{pages}{085317} (\bibinfo{year}{2003}).

\bibitem[{\citenamefont{Meier and Awschalom}(2004)}]{meier2004}
\bibinfo{author}{\bibfnamefont{F.}~\bibnamefont{Meier}} \bibnamefont{and}
  \bibinfo{author}{\bibfnamefont{D.~D.} \bibnamefont{Awschalom}},
  \bibinfo{journal}{cond-mat/0405342}  (\bibinfo{year}{2004}).

\bibitem[{\citenamefont{McKeever et~al.}(2004)}]{McKeever2004}
\bibinfo{author}{\bibfnamefont{J.}~\bibnamefont{McKeever}}
  \bibnamefont{et~al.}, \bibinfo{journal}{Science}
  \textbf{\bibinfo{volume}{303}}, \bibinfo{pages}{1992} (\bibinfo{year}{2004}).

\bibitem[{\citenamefont{Briegel et~al.}(1998)}]{Briegel98}
\bibinfo{author}{\bibfnamefont{H.}~\bibnamefont{Briegel}} \bibnamefont{et~al.},
  \bibinfo{journal}{Phys. Rev. Lett.} \textbf{\bibinfo{volume}{81}},
  \bibinfo{pages}{5932} (\bibinfo{year}{1998}).

\bibitem[{\citenamefont{Ekert}(1991)}]{ekert91}
\bibinfo{author}{\bibfnamefont{A.}~\bibnamefont{Ekert}},
  \bibinfo{journal}{Phys.\ Rev.\ Lett.} \textbf{\bibinfo{volume}{67}},
  \bibinfo{pages}{661} (\bibinfo{year}{1991}).

\bibitem[{\citenamefont{Lim et~al.}(2004)\citenamefont{Lim, Beige, and
  Kwek}}]{Lim2004}
\bibinfo{author}{\bibfnamefont{Y.}~\bibnamefont{Lim}},
  \bibinfo{author}{\bibfnamefont{A.}~\bibnamefont{Beige}}, \bibnamefont{and}
  \bibinfo{author}{\bibfnamefont{L.}~\bibnamefont{Kwek}}, \bibinfo{journal}{to
  appear}  (\bibinfo{year}{2004}).

\end{thebibliography}

\end{document}